\documentclass[prl,twocolumn,preprintnumbers,amsmath,amssymb, nofootinbib]{revtex4-1}
\usepackage{amsmath}
\usepackage{amssymb}
\usepackage{amsfonts}
\usepackage{eucal}
\usepackage{color}
\usepackage[active]{srcltx}%

\def\nn{\nonumber}

\newcommand{\ads}[1]{AdS$_{#1}$}
\newcommand{\cft}[1]{CFT$_{#1}$}

\newcommand{\sect}[1]{\noindent\textit{#1}.---\!}

\newcommand{\ba}{\begin{array}}
\newcommand{\ea}{\end{array}}
\newcommand{\be}{\begin{equation}}
\newcommand{\ee}{\end{equation}}
\newcommand{\bea}{\begin{eqnarray}}
\newcommand{\eea}{\end{eqnarray}}
\newcommand{\bse}{\begin{subequations}}
\newcommand{\ese}{\end{subequations}}
\newcommand{\bi}{\begin{itemize}}
\newcommand{\ei}{\end{itemize}}

\newcommand{\SO}{\mathrm{SO}}
\newcommand{\U}{\mathrm{U}}
\newcommand{\RR}{\mathbb{R}}

\definecolor{darkgreen}{rgb}{0,0.3,0}
\definecolor{darkblue}{rgb}{0,0,0.3}
\definecolor{darkred}{rgb}{0.7,0,0}

\begin{document}

\title{Extremal Black Holes and First Law of Thermodynamics}
\author{Maria Johnstone$^\dag$, M.M. Sheikh-Jabbari$^\natural$, Joan Sim\'on$^\dag$ and  Hossein Yavartanoo$^\ddag$}
\bigskip\medskip
\affiliation{$^\dag$ \textit{School of Mathematics and Maxwell Institute for Mathematical Sciences, King's Buildings, Edinburgh EH9 3JZ, United Kingdom}\\
$^\natural$ \textit{School of Physics, Institute for Research in Fundamental
Sciences (IPM), P.O.Box 19395-5531, Tehran, Iran}\\
\smallskip
$^\ddag$ {Department of Physics, Kyung Hee University, Seoul 130-701, Korea}}
\vfil

\setcounter{footnote}{0}

\begin{abstract}
\noindent
We study the low temperature expansion of the first law of thermodynamics for near-extremal black holes. We show that for extremal black holes with non-vanishing entropy, the leading order contribution {yields an expression for their extremal entropy in agreement with the entropy function result} and the Cardy formula for the entropy of a two dimensional chiral conformal field theory (CFT). When their entropy vanishes due to the vanishing of a {one-cycle} on the horizon, such leading contribution is always compatible with the first law satisfied by a BTZ black hole. These results are universal and consistent both with the presence of local AdS${}_2$ and AdS${}_3$ near horizon throats for extremal black holes and with the suggested quantum microscopic descriptions (AdS${}_2$/CFT${}_1$, Kerr/CFT and EVH/CFT).

\end{abstract}

\maketitle

\sect{Introduction} After the seminal works of Bardeen-Carter-Hawking \cite{BCH} and Bekenstein \cite{Bekenstein}, a deep connection between two seemingly independent branches of physics, gravity and thermodynamics,  was uncovered: {black holes satisfy the laws of thermodynamics \cite{BCH} and in particular the first law}
\be\label{1st-law-1}
T_{\text{H}}dS=dM-\sum_{i=1}^{n} \Omega^i dJ_i\,,
\ee
{where $S$ is their entropy, $M$ and $J_i$ their charges and $\Omega_i$ their conjugate chemical potentials.}
Even though the black hole temperature and entropy involve $\hbar$, this theorem relies on classical physics. Using a scalar quantum field $\Phi$ propagating in the black hole background, Hawking showed that black holes radiate \cite{Hawking}. In fact, the quantum state of $\Phi$ in the region outside the black hole horizon is a thermal density matrix $\hat{\varrho} \propto e^{-\beta\left(\omega-\sum_i n_i\Omega^i\right)}$ \cite{Hartle:1983ai,Frolov:1989jh}, with probe quantum numbers $\{\omega,\,n_i\}$ and $\beta=T_{\text{H}}^{-1}$, so that the temperature and chemical potentials of this radiation measured at infinity are those appearing in   \eqref{1st-law-1}.

The dependence on $\hbar$ and the existence of underlying statistical mechanical descriptions explaining the thermodynamic behaviour of systems with a large number of degrees of freedom gave rise to the belief that black holes hold the key to the quantum mechanical description of gravity. This expectation was realized in string theory where the macroscopic entropy of a certain class of (nearly) supersymmetric black holes was {reproduced by an explicit black hole microstate counting} \cite{Strominger:1996sh}.

This success is currently understood in terms of the AdS/CFT duality \cite{AdS/CFT}, so that all known such microscopic descriptions involve CFTs. More recently, it was proposed that generic \emph{extremal}, not necessarily supersymmetric, black holes admit a quantum mechanical description of their microstates in terms of CFTs \cite{Carlip,Ent-Funcn-Sen,Kerr/CFT,Hartman:2008pb}. This claim does not necessarily rely on string theory and is associated with the appearance of an AdS${}_2$ throat in the geometry close to the horizon of any extremal $(T_{\text{H}}=0$) black hole \cite{Kunduri:2007vf}. The goal is to extend this understanding to generic non-extremal black holes.

In this Letter we revisit the low temperature behaviour of near-extremal black holes. Using {some minimal inputs from black holes as solutions to gravity theories}, we identify the dominant contribution to the first law of thermodynamics in this limit. We show the latter is always satisfied by the near horizon geometries of these (near) extremal black holes, both for finite and vanishing  entropy, if the latter occurs through a vanishing one-cycle on the horizon. We comment on the relation and consistency of these results with the rigidity of these near-extremal geometries and the quantum mechanical proposals in the literature for the microscopic descriptions of the entropy of these black holes.

\sect{Black holes and the first law of thermodynamics} Black holes are solutions to classical theories of gravity, possibly coupled to matter, characterised by the existence of an event horizon.  When these theories have vacua, such as Minkowski or AdS,
where the notion of mass $M$ and conserved charges $J_i$, such as angular momentum or electric/magnetic charges, is properly defined \cite{Wald-review}, black holes carry non-trivial such charges.

The event horizon is a null hypersurface $\Gamma_{\text{H}}$ generated by a Killing vector field $\xi$ which is null on $\Gamma_{\text{H}}$. For regular black holes $\Gamma_{\text{H}}$ is smooth  and compact, and its area $A_{\text{H}}$ determines the entropy $S$, for Einstein gravity theory $S=\frac{A_{\text{H}}}{4G_N}$ \cite{Hawking:1974sw}. The remaining thermodynamical quantities in \eqref{1st-law-1} are determined by local horizon properties \cite{Wald-review}
\begin{subequations}
\begin{align}
  T_H &= \frac{\kappa}{2\pi}\,, \quad \xi\cdot\xi\big|_{\Gamma_{\text{H}}}=0\,, \label{horizon-chargesa}\\
\Omega^i&=\xi\cdot A^{(i)}\big|_{\Gamma_{\text{H}}}\,,\quad i=m+1,\cdots, n\,, \label{horizon-chargesb}
\end{align}
\end{subequations}
where $\kappa^2=-D^\mu\xi^\nu D_\mu\xi_\nu/2$ is the horizon surface gravity. Notice we implicitly split $\Omega^i$ into angular velocities $(i=1,\dots, m)$ and chemical potentials $(i=m+1, \dots, n)$ associated with electric fields $A^{(i)}$ \cite{dipole-charge}.

To gain some geometric intuition consider, for ease of notation, 4d/5d black holes with $\RR\times \U(1)^{d-3}$ invariant metrics which in adapted coordinates look like
\begin{equation}
\hspace*{-1mm}  ds^2_{\text{bh}} = -N^2dt^2 + \frac{dr^2}{g^{rr}} + g_{\theta\theta}d\theta^2 + \gamma_{ab}(d\phi^a - \mathcal{B}^adt)(d\phi^b - \mathcal{B}^bdt)\,,
\label{eq:bh}
\end{equation}
where all components are functions of $r,\theta$. The event horizon is at $r=r_+$, the largest root of $g^{rr}(r_+)=0$. $\xi=\partial_t + \sum_{i=1}^{d-3}\Omega^a\partial_{\phi^a}$ is null on this surface. It determines the horizon angular velocities {$\Omega^a=\mathcal{B}^a(r_+)$}, using \eqref{horizon-chargesa}, and the rest of the $\Omega^i$'s by the gauge field values at $r_+$ \eqref{horizon-chargesb}. In the neighbourhood of the horizon, one can always introduce Gaussian null coordinates  $(v,\rho,x^a)$ in which
\begin{equation}
\label{eq:gaussian}
 \hspace*{-2mm} ds^2 = \rho^2F(\rho,x)dv^2+2dvd\rho+2\rho h_I(\rho,x)dvdx^I+\gamma(\rho,x), \hspace{-1mm}
\end{equation}
where $\xi=\partial_v$, $\rho=0$ is the horizon and $x^I$ describe the spatial compact horizon cross-section with metric {$\gamma=\gamma_{IJ}(\rho,x^I)dx^Idx^J$}.

\sect{Finite entropy near-extremal systems}
Given a thermodynamical system with $n+1$ charges, the limit of vanishing temperature defines an $n$-dimensional extremal surface (ES) characterised by the relation $M_e=M_e(J_i)$, compatible with \eqref{1st-law-1}:
\be\label{M-J-extremal}
dM_e=\sum_{i=1}^n\Omega^i_e dJ_i \,,\quad \Omega^i_e=\frac{\partial M}{\partial J_i}\bigg|_{{}_{\mathrm{ES}}}\,.
\ee
We study the first non-trivial correction due to a small but non-zero temperature for a near-extremal black hole with non-vanishing extremal entropy $S_e\neq 0$, which is achieved by a deformation $(M_e+\delta M,\,J_i)$ orthogonal to the ES. Assuming {\it analyticity} of all thermodynamical quantities, we define this deformation as
\be\label{eq:delta}
T_{\text{H}}=T^\prime\mid_{{}_{\mathrm{ES}}}\epsilon,\ \  \Omega^i=\Omega^i_e +\Omega^{i\prime}\mid_{{}_{\mathrm{ES}}}\epsilon\, ,
\ee
where \textit{prime} denotes differentiation with respect to the deviation from extremality, i.e. orthogonal motion away from the ES. The question we explore is which physical excitations are relevant at first order in $\epsilon$ for {regular} black holes. To this end we first show that $\delta M$ is second order in $\epsilon$
\be\label{extremality-deviation}
dM-\sum_i\Omega^i_e dJ_i\equiv d\left(\delta M\right) \propto \epsilon^2\,.
\ee
To prove this, recall that  $M_e=M_e(J_i)$ corresponds to an extremal black hole with equal outer $(r_+)$ and inner $(r_-)$ horizon radii, and that $T_{\text{H}}\sim r_+-r_-\sim \epsilon$ for near-extremal black holes. Equivalently,  if $r_e$ is the extremal black hole horizon radius, turning on temperature corresponds to a deformation in black hole parameter space satisfying $r_\pm = r_e\pm \Delta\epsilon$, where {$\Delta\sim \epsilon^0$ is  defined on the ES}. Then, let $f(r; M,J_i)=0$ be the equation specifying the location of the horizon(s). At an extremal point $f(r, M_e,J_i)= C (r-r_e)^2$. Turning on $\delta M$, and relying on analyticity,
$$
f(r; M_e+\delta M, J_i)= \delta A+ \delta B (r-r_e)+ (C+\delta C) (r-r_e)^2+ \dots,
$$
with $\delta A$ and $\delta B$ proportional to $\delta M$. Requiring $f$ to have roots at $r_\pm$ determines $ \delta A=-C \epsilon^2\Delta^2,\quad \delta B= {\cal O}(\epsilon^2)$ and hence \eqref{extremality-deviation} follows.

The first law \eqref{1st-law-1} at first order in $\epsilon$ then reduces to \cite{Azeyanagi:2009wf}
\be\label{1st-law-extremal-expanded}
dS_e=\sum_{i=1}^n k^i{dJ_i}\,,\quad  k^i=k^i(J_k)= -\frac{\Omega^{i\prime}}{T^\prime}\bigg|_{{}_{\mathrm{ES}}}\,.
\ee
Note that, despite considering the near-extremal limit \eqref{eq:delta}, all quantities in \eqref{1st-law-extremal-expanded}, including the entropy, are evaluated on the ES and are independent of the deformation $\epsilon$. Thus, \eqref{extremality-deviation} implies that the low temperature expansion of the first law for regular black holes in the leading order  yields information on the entropy of the extremal black holes and not on excitations above them. The latter may arise at second or higher orders in the expansion of the first law \cite{Castro:2009jf}.

In the above we merely used thermodynamic considerations and the only crucial input from black holes is \eqref{extremality-deviation}. Below, we will interpret this condition, but now we want to prove that regular black holes {\it do} satisfy \eqref{extremality-deviation}. We show this using Sen's entropy function formalism \cite{Ent-Funcn-Sen}, which relies on the near horizon geometry of regular extremal black holes.

It was {proved} in \cite{Kunduri:2007vf} that any regular extremal black hole with $\RR\times \U(1)^{d-3}$ isometry group has an {\it on-shell near horizon} geometry of the form \cite{end-note}
\begin{equation}
\label{eq:gnh}
ds^2=\Gamma(\theta) \bar{g} + d\theta^2 +\gamma_{ab}(\theta) (d\varphi^a +e^a \rho dt)(d\varphi^b +e^b \rho dt)
\end{equation}
with $\bar{g}=-\rho^2 d\tau^2 +d\rho^2/\rho^2$. While \eqref{eq:gnh} is proved to hold for any 4d or 5d classical theory of Einstein gravity coupled to an arbitrary number of Abelian Maxwell-type gauge fields and neutral scalar fields with lagrangian density ${\cal L}$, it is {known} to hold for more general extremal black holes.

This solution is invariant under $\SO(2,1)\times \U(1)^{n}$. While $\SO(2,1)$ is geometrically realized as the isometries of the AdS${}_2$ $\bar{g}$,  $n-d+3$ of the $\U(1)$ symmetries are associated with non-geometrical Abelian gauge fields.

{The presence of} $\SO(2,1)$ allows the entropy function formalism to show that solving the equations of motion (eom) for backgrounds \eqref{eq:gnh} is equivalent to extremising the entropy function ${\cal E}$
\begin{equation}
{\cal E} = 2\pi\left(\sum_{i=1}^n e^iJ_i - f(\Gamma,\gamma_{ab},J_i)\right)
\label{eq:ef}
\end{equation}
where $f = \int_{\Gamma_{\text{H}}} \sqrt{-\text{det}\ g}\, {\cal L}$. It also asserts that the on-shell evaluation of ${\cal E}$ equals the entropy of the extremal black hole, i.e. ${\cal E}_{\text{eom}}=S_e$. Computing the on-shell variation of \eqref{eq:ef} yields
\begin{equation}
  dS_e = 2\pi \sum_{i=1}^n e^i\ dJ_i
\end{equation}
which matches \eqref{1st-law-extremal-expanded} if $k^i ={2\pi e^i}$. This identification is manifest when studying the near horizon limit $r=r_+ + \epsilon \rho$ in metrics \eqref{eq:bh} in a near-extremal regime \cite{{Ent-Funcn-Sen},Azeyanagi:2009wf,Chow:2008dp}
\begin{equation}\label{NH-ext-scaling}
  t=\frac{\tau}{2\pi T^\prime\epsilon},\quad \varphi^i = \phi^i -\Omega^i_e\frac{\tau}{2\pi T^\prime\epsilon}, \quad \epsilon\to 0
\end{equation}
where $\phi^i, \varphi^i$ denote both angular isometry directions and gauge fields. The expansions $X^i = \Omega^i_{\text{e}} + \Omega^{i\prime} \epsilon\rho$, where $X^i=\mathcal{B}^i$ for $i=1,\cdots d-3$ in \eqref{eq:bh} and $X^i$ equals the gauge fields $A^{(i)}$ for $i=d-2,\cdots, n$, confirms our assertion and establishes that {\it on-shell} near horizon geometries of extremal black holes {\it always} satisfy \eqref{1st-law-extremal-expanded}.

A similar conclusion can be reached by probing the extremal black hole background by a quantum field $\Phi$. The near-extremal limit \eqref{eq:delta} requires $\omega=\sum_i \Omega_e^in_i + {\cal O}(\epsilon^\alpha)$ to prevent the vanishing of the probability density $\hat\varrho$. This is the analogue of the black hole extremality condition \eqref{M-J-extremal} and spectrum of excitations \eqref{extremality-deviation} for the scalar probe. In the limit $\epsilon\to 0$, $\hat{\varrho}\propto e^{\sum_i k_i(J_k) n_i}$ reproduces the chemical potentials in \eqref{1st-law-extremal-expanded}.
{In fact, \eqref{extremality-deviation} guarantees that {\it gravitational} perturbations have $\alpha=2$, hence they do not add extra terms to $\hat{\varrho}$ and consequently to the first law, at first order, in the limit \eqref{eq:delta}.}

\sect{Vanishing entropy near-extremal systems}
Systems satisfying $S_e\sim \epsilon^p, \ p>0$ as $T_{\text{H}}\sim\epsilon\to 0$ do not obey our previous analysis. Here we consider black holes with $S_e/T_{\text{H}}$ {\it finite}, i.e. $p=1$. Since black holes in this category have vanishing horizon area, we refer to them as extremal vanishing horizon (EVH) ones. In parameter space, they exist at the co-dimension $k$ ($k\geq 2$) {\it EVH surface} defined by the intersection of the $T_{\text{H}}=S_e=0$ surfaces.

Inspired by the finiteness of $S_e/T_{\text{H}}$ and the Cardy-like growth in the density of states of a 2d CFT,
we want to identify a set of sufficient conditions under which $p=1$ EVH black holes allow a first law of thermodynamics compatible with a 2d CFT microscopic description. For this purpose, we  {\it assume} vanishing of the entropy is due to the vanishing of a single {{\it one-cycle}}. This requires the existence of a vanishing eigenvalue in the near horizon metric $\gamma_{ab}$ in \eqref{eq:gnh}. Let this eigenvalue be along the $\partial_\varphi$ direction. Then, close to the horizon $r\to\epsilon r,\ \epsilon\to 0$, the length of this one-cycle must scale like $\gamma_{\varphi\varphi}\sim \gamma(\theta)\,r^2$. Smoothness requires $\gamma(\theta) = \Gamma (\theta)$, as in \eqref{eq:gnh}, giving rise to
\begin{equation}
\label{eq:adst}
  ds^2 = \Gamma(\theta)\left(\epsilon^2 r^2(-dt^2 + d\varphi^2)+\frac{dr^2}{r^2}\right) + ds^2_\perp
\end{equation}
as the {\it near horizon} geometry of such EVH black hole. Notice the isometry of the first factor is locally enhanced to $\SO(2,2)$. Requiring  a smooth  geometry at $r=0$ and $ds^2_\perp$ to be $\SO(2,2)$ invariant implies that $ds^2_\perp$ is $r$ independent. These statements were proved in 4d Einstein-Maxwell-dilaton theories \cite{KKEVH} and have been verified in several 4d and 5d studies \cite{Massless-BTZ,KKEVH,Rot-AdS-BH}. We expect them to hold for any such EVH black hole. We will assume them hereafter.

In view of \eqref{eq:adst}, there exist two inequivalent near-horizon geometries : the  null self-dual orbifold of \ads{3} \cite{SDorb} after redefining $t + \varphi\to \left(t + \varphi\right)/\epsilon^2$ keeping $t-\varphi$ fixed and the {\it pinching} \ads{3} orbifold \cite{Massless-BTZ} after $t, \varphi\to t/\epsilon, \varphi/\epsilon$. The former reduces the isometry to $\SO(2,1)\times \U(1)$ and does not allow for dynamical excitations \cite{Massless-BTZ}. Searching for non-trivial excitations, responsible for $d\delta M\sim TdS_e\sim \epsilon^2$ in \eqref{extremality-deviation} when we turn on the temperature, we will focus on the second option.

Note that the 3d metric in \eqref{eq:adst} can be obtained by taking the near horizon limit $(r\to\epsilon r$) of metrics \eqref{eq:bh}
\begin{equation}
 \label{eq:nhevh}
   t=\frac{\tau}{\epsilon},\ \ \varphi_i = \phi_i -\Omega^{(0)}_i\frac{\tau}{\epsilon}, \ \ \chi = \epsilon\ \varphi\,, \quad \epsilon\to 0\,, \
 \end{equation}
where $\phi_i$, $i=1,\cdots,n-1$ parameterize the non-singular $\U(1)$ symmetries of the original EVH black hole solution. These include both the isometries of $ds^2_\perp$ and the Abelian $\U(1)$ symmetries due to the gauge fields of the solution. The resulting geometry includes the {\it pinching} \ads{3} orbifold \cite{Massless-BTZ}, due to $\chi\sim \chi+2\pi\epsilon$ and it is written in a co-rotating frame since $\Omega^{(0)}_i$ are the horizon angular velocities or electric potentials at the EVH point.

Having identified the relevant EVH geometry, we now turn to thermodynamics and switch on temperature and entropy
\be\label{EVH-expansion-2}
T_H=\epsilon T^{(1)}\,,\ S_H=\epsilon S^{(1)}\,,
\ee
along the $k$ orthogonal directions in parameter space that infinitesimally move away from the EVH surface. By construction, $dS_H$ is a one-form belonging to the space of EVH orthogonal deformations. In gravity, these deformations will correspond to near-EVH black holes. By analyticity,  any thermodynamical quantity $Z$ will allow an expansion $Z = Z^{\text{EVH}} + \sum_{n>0}\epsilon^n Z^{(n)}$, with $dZ^{\text{EVH}}=0$. Besides \eqref{EVH-expansion-2}, we have
\begin{align}\label{EVH-expansion-1}
dM&=\epsilon dM^{(1)}+\epsilon^2 dM^{(2)}+\cdots\,,\cr
dJ_i&=\epsilon dJ_i^{(1)}+\epsilon^2 dJ_i^{(2)}+\cdots\,,\\
\Omega_i&=\Omega_i^{(0)}+\epsilon \Omega_i^{(1)}+\cdots\,.\nn
\end{align}

Since $T_{\text{H}}dS \sim \epsilon^2$, the $\Omega_i$  in the near-EVH regime relevant to our leading order analysis are of the form
\be\label{Omega-EVH}
\Omega_i^{\text{near-EVH}}=\left\{\begin{array}{cc} \Omega_a^{(0)}&\qquad a=k-1,\cdots, n\,,\\
\epsilon\Omega_\alpha^{(1)} &\qquad \alpha=1,\cdots, k-2\,,\\
\Omega_\varphi&\, \text{for pinching direction}\ \varphi\,.\end{array}\right.
\ee
We singled out the angular velocity along the pinching direction, because even though it vanishes at the EVH point, its value in the near-EVH regime can be non-vanishing. This is because $\Omega_\varphi\sim d\varphi/dt$ and as discussed previously, both coordinates scale in the same way. Noting the coordinate  scalings \eqref{eq:nhevh}, a similar argument leads to the scaling of $\Omega_a$ and $\Omega_\alpha$ given in \eqref{Omega-EVH}. There is no second order contribution, because at this order the only non-trivial contribution would come from $dJ_i^{(0)}$ which vanishes, by construction.

Using \eqref{extremality-deviation}, i.e. $d\delta M \equiv \epsilon^2\left(dM^{(2)}-\sum_a\Omega^{(0)}_a dJ_a^{(2)}\right)$, and
plugging \eqref{EVH-expansion-2}, \eqref{EVH-expansion-1} and \eqref{Omega-EVH} into the first law, one obtains
\begin{equation}
T^{(1)} dS^{(1)}= d\delta M -\Omega_\varphi dJ^{(2)}_\varphi -\sum_\alpha \Omega^{(1)}_\alpha dJ^{(1)}_\alpha\,.
\label{second-order-EVH}
\end{equation}
As we will show below, \eqref{second-order-EVH} is indeed the first law for a BTZ black hole \cite{BTZ}.

Consider the subset of $\Omega^{(1)}_\alpha$ that are geometrically realised. One can always work in a co-rotating frame by performing the coordinate transformation
\be\label{co-rotating}
\phi_\alpha\ \to\ \varphi_\alpha=\phi_\alpha-A_\alpha \frac{\tau}{\epsilon}-B_\alpha \frac{\chi}{\epsilon}\,,
\ee
where $A_{\alpha}$ and $B_\alpha$ are first order in $\epsilon$, extending the transformation \eqref{eq:nhevh} to the near-EVH regime. Equivalently, these angular velocities must satisfy the identity
\be\label{Omega-alpha}
\Omega^{(1)}_\alpha= A^{(1)}_\alpha+B^{(1)}_\alpha\Omega_\varphi\,.
\ee
For those $\Omega_\alpha$ that are not geometrical, the corresponding transformation is a $\U(1)$ gauge transformation. Either way, recalling  symmetries of the near-EVH solution, \eqref{co-rotating} is just a large gauge or coordinate transformation by an element in the  $\SO(2,2)\times \U(1)^{k-1}$ symmetry group.

Since $\Omega_\alpha$ and $J_\alpha$ are first order deformations belonging to a $k-2$ dimensional space, they must be linearly related as $\Omega_\alpha=M_{\alpha\beta}J_\beta$, where only the symmetric part of the matrix $M_{\alpha\beta}$ is relevant for the first law \eqref{1st-law-1}. It follows that
\be
\begin{split} &
d(\sum_\alpha A^{(1)}_\alpha J^{(1)}_\alpha)=2\sum_\alpha A^{(1)}_\alpha dJ^{(1)}_\alpha\,,
\cr &
d (\sum_\alpha B^{(1)}_\alpha J^{(1)}_\alpha)=2\sum_\alpha B^{(1)}_\alpha dJ^{(1)}_\alpha\,.
\end{split}
\label{eq:smarr0}
\ee

Plugging \eqref{Omega-alpha} and \eqref{eq:smarr0} into \eqref{second-order-EVH}, it simplifies to
\be\label{BTZ-first-law}
T_{\text{BTZ}}\ dS_{\text{BTZ}}=d M_{\text{BTZ}}-\Omega_{\text{BTZ}} d J_{\text{BTZ}}\,{,}
\ee
where we defined
\begin{equation}\label{BTZ-mass-J}
\begin{split}
T^{(1)}&=T_{\text{BTZ}}\,,\quad S^{(1)}=S_{\text{BTZ}}\,,\quad \Omega_\varphi=\Omega_{\text{BTZ}}\,, \\
M_{\text{BTZ}}&= M^{(2)}-\sum_a \Omega_a^{(0)}J_a^{(2)}-\frac12\sum_\alpha A^{(1)}_\alpha J^{(1)}_\alpha \,,\\
J_{\text{BTZ}} &= J^{(2)}_\varphi+\frac12\sum_\alpha B^{(1)}_\alpha  J^{(1)}_\alpha\,.
\end{split}
\end{equation}
The BTZ subscript is to stress the fact that \eqref{BTZ-first-law} has the same form as the first law for a
BTZ black hole \cite{BTZ}. This reinforces the expectation that near-EVH black holes have a universal near horizon geometry given by a pinching BTZ, as explicitly checked for several near-EVH black holes, e.g. see \cite{KKEVH,Rot-AdS-BH}. This is consistent with the idea that near-EVH deformations in parameter space  correspond to large gauge transformations in gravity.

This result can be reproduced by a scalar field $\Phi$ probing the EVH black hole. Consider the Fourier decomposition
$\Phi \sim e^{-i\omega t}\,e^{i\sum_j m_j\phi_j}\,F(r,\theta_k)$. Using \eqref{eq:nhevh} and \eqref{co-rotating}, one identifies the probe quantum numbers in the IR near horizon geometry as $\omega = \epsilon\hat \omega +\Omega_i\,n_i + A_\alpha n_\alpha$, $m_\varphi= \epsilon\hat n_\varphi - B_\alpha n_\alpha$. Rewriting the density matrix $\hat \rho$ in terms of these IR quantum numbers, one derives $\hat\rho \propto e^{-\beta_{\text{BTZ}}\left(\hat{\omega}-\Omega_{\text{BTZ}}\hat{n}_\varphi\right)}$, where we used \eqref{EVH-expansion-2} and \eqref{Omega-alpha}. Thus, the $\Phi$ quantum state is the same as one outside of a BTZ black hole, in agreement with \eqref{BTZ-first-law}.

\sect{Comments on microscopic descriptions}
We conclude by relating the main physical inputs leading to \eqref{1st-law-extremal-expanded} and \eqref{BTZ-first-law} to the proposed microscopic descriptions in the literature. First, condition \eqref{extremality-deviation} is a measure of the rigidity of the near horizon geometries associated to {near} extremal black holes. In our thermodynamic analysis, it determined the absence of physical excitations at first order in $\epsilon$.
This agrees with the absence of near horizon AdS${}_2$ excitations preserving both AdS${}_2$ boundaries \cite{Maldacena:1998uz}, the dual formulation of these systems in terms of conformal quantum mechanics with no normalizable excitations \cite{Sen:2008vm} and the absence of normalizable gravitational excitations of the near horizon of extremal Kerr \cite{Amsel:2009ev}. Second,
it is worth stressing that \eqref{1st-law-extremal-expanded} is always compatible with a Cardy-like density of states growth at temperature $T=k(J)^{-1}$ whenever $n=1$. The latter does not uniquely select the existence of a 2d chiral CFT microscopic interpretation, but it is certainly consistent with it. Indeed, the central charge of such CFT would equal $c=3S_e/(\pi^2k)$. This is in agreement with the Kerr/CFT proposal \cite{Kerr/CFT}, which identifies a chiral Virasoro algebra with this central charge as the asymptotic symmetry group of the {\it same} near horizon geometries \eqref{eq:gnh} with some prescribed boundary conditions.

{It is known, via the AdS$_3$/CFT$_2$ duality \cite{AdS/CFT}, that a generic BTZ black hole corresponds to
a thermal mixed state in the dual 2d non-chiral CFT{,} and an extremal BTZ to a mixed state at zero temperature. It was shown that the near horizon limit of extremal BTZ corresponds to the Discrete Light-Cone Quantisation (DLCQ) of the dual CFT$_2$ which has only one dynamical chiral sector, leading to a 2d chiral CFT  \cite{DLCQ-paper}.} This explains why the chiral sector associated with AdS${}_2$ excitations is frozen (if suitable boundary conditions are used) and identifies the chiral Kerr/CFT sector with the second chiral sector of the original 2d CFT.

This also ties nicely with our result \eqref{BTZ-first-law} for EVH black holes. As explained in \cite{Massless-BTZ}, the AdS${}_3$ pinching orbifold corresponds to a limit in the non-chiral 2d CFT in which both chiral sectors decouple. To keep some physical degrees of freedom, one can accompany the near-EVH limit with a scaling of Newton's constant $G_N\to 0$, so that the entropy of the system remains finite. This is accomplished by large $N$ limits in CFTs with gravity duals \cite{KKEVH,Massless-BTZ}. The thermodynamic analysis presented here gives strong evidence that the same structure exists for any EVH black hole {where the vanishing of the horizon area is due to a vanishing one-cycle}, as conjectured in \cite{KKEVH} as a specific case of the \ads{3}/\cft{2} correspondence \cite{AdS/CFT}.
\acknowledgments

We would like to thank Micha Berkooz for useful comments. MMSHJ, JS and HY would like to thank CQUeST for hospitality during the {\it Quantum Aspects of Black Holes} workshop. The work of MJ and JS was partially supported by the Engineering and Physical Sciences Research Council (EPSRC) [grant number EP/G007985/1] and the Science and Technology Facilities Council (STFC) [grant number ST/J000329/1].  The work of HY was supported by the National Research Foundation of Korea Grant funded by the Korean Government (NRF-2011- 0023230).

\end{document}